\documentclass[aps,prl,twocolumn,showpacs,superscriptaddress]{revtex4-1}
\usepackage{graphicx}
\usepackage{amssymb}
\usepackage{amsmath}
\usepackage{here}
\usepackage{color}
\usepackage{soul}

\begin{document}

\title{Spin-lattice coupling in a ferrimagnetic spinel:
The exotic $H$--$T$ phase diagram of MnCr$_2$S$_4$ up to 110~T}

\author{A. Miyata}

\affiliation{Laboratoire National des Champs Magn\'{e}tiques Intenses, (LNCMI-EMFL), CNRS-UGA-UPS-INSA, 31400 Toulouse, France}

\author{H. Suwa}

\affiliation{Department of Physics and Astronomy, The University of Tennessee, Knoxville, TN 37996, USA}

\affiliation{Department of Physics, The University of Tokyo, Tokyo 113-0033, Japan}

\author{T. Nomura}

\affiliation{Hochfeld-Magnetlabor Dresden (HLD-EMFL) and W\"urzburg-Dresden Cluster of Excellence ct.qmat, Helmholtz-Zentrum Dresden-Rossendorf, 01328 Dresden, Germany}

\author{L. Prodan}

\affiliation{Institute of Applied Physics, MD 2028, Chisinau, R. Moldova}

\author{V. Felea}

\affiliation{Hochfeld-Magnetlabor Dresden (HLD-EMFL) and W\"urzburg-Dresden Cluster of Excellence ct.qmat, Helmholtz-Zentrum Dresden-Rossendorf, 01328 Dresden, Germany}

\affiliation{Institute of Applied Physics, MD 2028, Chisinau, R. Moldova}

\affiliation{Institut f\"ur Festk\"orper- und Materialphysik, TU Dresden, 01069 Dresden, Germany}

\author{Y. Skourski}

\affiliation{Hochfeld-Magnetlabor Dresden (HLD-EMFL) and W\"urzburg-Dresden Cluster of Excellence ct.qmat, Helmholtz-Zentrum Dresden-Rossendorf, 01328 Dresden, Germany}

\author{J. Deisenhofer}

\affiliation{Experimental Physics 5, Center for Electronic Correlations and Magnetism, Institute of Physics, University of Augsburg, 86159, Augsburg, Germany}

\author{H.-A. Krug von Nidda}

\affiliation{Experimental Physics 5, Center for Electronic Correlations and Magnetism, Institute of Physics, University of Augsburg, 86159, Augsburg, Germany}

\author{O. Portugall}

\affiliation{Laboratoire National des Champs Magn\'{e}tiques Intenses, (LNCMI-EMFL), CNRS-UGA-UPS-INSA, 31400 Toulouse, France}

\author{S. Zherlitsyn}

\affiliation{Hochfeld-Magnetlabor Dresden (HLD-EMFL) and W\"urzburg-Dresden Cluster of Excellence ct.qmat, Helmholtz-Zentrum Dresden-Rossendorf, 01328 Dresden, Germany}

\author{V. Tsurkan}

\affiliation{Institute of Applied Physics, MD 2028, Chisinau, R. Moldova}

\affiliation{Experimental Physics 5, Center for Electronic Correlations and Magnetism, Institute of Physics, University of Augsburg, 86159, Augsburg, Germany}

\author{J. Wosnitza}

\affiliation{Hochfeld-Magnetlabor Dresden (HLD-EMFL) and W\"urzburg-Dresden Cluster of Excellence ct.qmat, Helmholtz-Zentrum Dresden-Rossendorf, 01328 Dresden, Germany}

\affiliation{Institut f\"ur Festk\"orper- und Materialphysik, TU Dresden, 01069 Dresden, Germany}

\author{A. Loidl}

\affiliation{Experimental Physics 5, Center for Electronic Correlations and Magnetism, Institute of Physics, University of Augsburg, 86159, Augsburg, Germany}

\date{}

\begin{abstract}

  In antiferromagnets, the interplay of spin frustration and spin-lattice coupling has been extensively studied as the source of complex spin patterns and exotic magnetism. Here, we demonstrate that, although neglected in the past, the spin-lattice coupling is essential to \textit{ferrimagnetic} spinels as well. We performed ultrahigh-field magnetization measurements up to 110~T on a Yafet-Kittel ferrimagnetic spinel, MnCr$_2$S$_4$, which was complemented by measurements of magnetostriction and sound velocities up to 60~T. Classical Monte Carlo calculations were performed to identify the complex high-field spin structures. Our minimal model incorporating spin-lattice coupling accounts for the experimental results and corroborates the \textit{complete} phase diagram, including two new high-field phase transitions at 75 and 85~T. Magnetoelastic coupling induces striking effects: An extremely robust magnetization plateau is embedded between two unconventional spin-asymmetric phases. Ferrimagnetic spinels provide a new platform to study asymmetric and multiferroic phases stabilized by spin-lattice coupling.

\end{abstract}

\maketitle

\section{Introduction}
Ferrimagnets, thanks to their finite spontaneous magnetization, have been widely used in engineering and technology, for example, as strong permanent magnets, optical isolators, and circulators for optical communications. Unlike ferromagnets, by tuning the antiferromagnetic exchanges in ferrimagnets, additional functionalities, e.g., ferroelectricity and skyrmion controllability \cite{Kim03, Sek12}, may emerge as a consequence of a noncollinear spin structure. The realization of such multifunctionality opens the door for next-generation technologies.

Noncollinear ferrimagnets have been well studied, for instance, in spinels, $AB_2X_4$, where the $A$ and $B$ cations form a bipartite diamond lattice and a pyrochlore lattice, respectively, with $X$ = O, S, or Se. The key essence is the competition (or frustration) of magnetic exchanges within or between the two $A$ and $B$ lattices ($J_{A\text{-}A}$, $J_{B\text{-}B}$, and $J_{A\text{-}B}$). Yafet and Kittel (YK) proposed a model for triangular-structure ground states, where $J_{A\text{-}A} \sim J_{A\text{-}B}$ \cite{Yaf52}, while Lyons, Kaplan, Dwight, and Menyuk (LKDM) proposed another model for conical-structure ground states, where $J_{B\text{-}B} \sim J_{A\text{-}B}$ \cite{Lyo62}. Both noncollinear ground states have been widely observed in ferrimagnetic spinels \cite{Coe87, Kap07}.

To realize unconventional ferrimagnetic structures beyond both the YK and the LKDM models, one can consider that spontaneous lattice deformation will modulate these main antiferromagnetic exchanges, i.e., through a spin-lattice coupling mechanism \cite{Pen04}. Such lattice deformation has been expected in strongly frustrated antiferromagnets to lift the macroscopic degeneracy (or lower the free energy of the system), resulting in unconventional magneto-structural phases \cite{Yam00, Tch02}.  On the other hand, this kind of spin-lattice coupling mechanism has not been taken into account in previous theoretical works on ferrimagnetic spinels, because they have been considered to be less frustrated. Here, our experimental and theoretical studies demonstrate that the spin-lattice coupling is essential in ferrimagnetic spinels as well.

MnCr$_2$S$_4$  ($A$ = Mn$^{2+}$ with spin \emph{S} = 5/2, and $B$ = Cr$^{3+}$ with \emph{S} = 3/2 in the spinel structure, see Fig.~1a) is a representative material for the YK model, where \emph{J}$_\text{Mn-Mn}$ $\sim$ \emph{J}$_\text{Mn-Cr}$ \cite{Lot56, Men62, Men65, Lot68, Tsu03}. 
The interaction between the Cr spins, \emph{J}$_\text{Cr-Cr}$, is ferromagnetic and much stronger than the other spin interactions, which has been evidenced by two consecutive magnetic phase transitions at $T_\text{C}\approx$ 65~K (ferromagnetic order of Cr spins) and at $T_\text{YK}\approx$ 5 K (triangular-like magnetic order of Cr and Mn spins) \cite{Lot56, Tsu03}. 
Recent high-field ultrasound and magnetization experiments on MnCr$_2$S$_4$ up to 60~T revealed a complex phase diagram and novel magneto-structural phases \cite{Tsu17}. 
An extremely robust magnetization plateau at 6~$\mu$$_\text{B}$/f.u. between 25 and 50~T was observed. 
In analogy to the models proposed by Matsuda and Tsuneto \cite{Mat70} and Liu and Fisher \cite{Liu73}, Tsurkan \textit{et al}. proposed that the magnetic orders of the Mn spin below and above the magnetization plateau should be interpreted as spin supersolid-like phases \cite{Tsu17}.

Furthermore, it has been revealed very recently that the supersolid and superfluid-like phases of MnCr$_2$S$_4$ are ferroelectric, i.e., multiferroic \cite{Ruf18}.
It is, thus, important to clarify the magnetic structures of MnCr$_2$S$_4$ in the sense of multiferroicity as well.
The understanding of the phases could pave the way to exploit \textit{ferrimagnetic} multiferroicity, where finite magnetization moments can be utilized, e.g., the electric-field reversal of ferrimagnetic moments \cite{Tok12}.

In this paper, we report on the magnetization process of MnCr$_2$S$_4$ in magnetic fields up to 110~T and show the \textit{complete} phase diagram of the YK-type ferrimagnetic spinel. 
The magnetization experiments were complemented by magnetostriction and ultrasound measurements up to 60~T, which evidence strong spin-lattice couplings.
We propose a minimal Hamiltonian for describing the magnetic and structural properties of MnCr$_2$S$_4$, incorporating spin-lattice coupling between the Mn and Cr spins.
Using the classical Monte Carlo method we compare the theoretical and experimental data regarding both the \textit{spin} and \textit{lattice} degrees of freedom. 
Our analysis evidences that the spin-lattice coupling stabilizes the magnetization plateau and unconventional phases of asymmetric magnetic structures.
The obtained phase diagram is remarkably symmetric with respect to the center of the magnetization plateau ($\sim$40~T), where the external field perfectly cancels out the internal exchange field acting on the Mn spins. 

\section{Methods}

Magnetization measurements were performed at the LNCMI-EMFL in Toulouse in pulsed magnetic fields up to 110~T along ${\bf H} \parallel [110]$ using a single-turn-coil technique \cite{Por99}.
The magnetization was measured using the induction method with a compensated pair of coils as described in ref. \cite{Tak12}.

The optical fibre Bragg grating (FBG) method was used to measure the magnetostriction \cite{Daou10} up to 60 T at the HLD-EMFL in Dresden.
The relative length change $\Delta L/L$ is obtained from the shift of the Bragg wavelength of the FBG.
The magnetostriction was measured along ${\bf H} \parallel [110]$.

Ultrasound measurements were performed using the standard pulse-echo technique to investigate the elastic properties \cite{Zherlitsyn14} up to 60 T at the HLD-EMFL in Dresden. 
The longitudinal $(c_{11}+2c_{12}+4c_{44})/3$ mode was studied with the alignment of ${\bf H} \parallel {\bf k} \parallel {\bf u} \parallel [111]$, where ${\bf k}$ and ${\bf u}$ are propagation and polarization vector, respectively.
Polyvinylidene fluoride (PVDF) film transducers were used to excite and to detect ultrasonic waves at the frequency of 65 MHz.

Classical (Markov chain) MC simulations were performed for the system described by eq.~(1). The spin and lattice configurations were sampled from the Boltzmann distribution at finite temperatures. The number of sites and bonds in the simulations are 5,184 and 34,560, respectively. We confirmed that the results for 5,184 and 1,536 sites are consistent within error bars: the size effect is negligible and the present results are representative for the thermodynamic limit. We adopted the overrelaxation technique both for the spin and lattice degrees of freedom, which enables rejection-free updates with the energy constant. MC updates with the Metropolis algorithm were combined for thermalization. We repeated the whole process consisting of one Metropolis and five overrelaxation steps. We took the average over $2^{21}$ MC samples after $2^{16}$ thermalization steps, which are much longer than the autocorrelation time. The error bars are comparable to or smaller than the linewidths in the present figures.

\begin{figure*}[t]
----
\centering
\includegraphics[angle=0,width=0.8\textwidth]{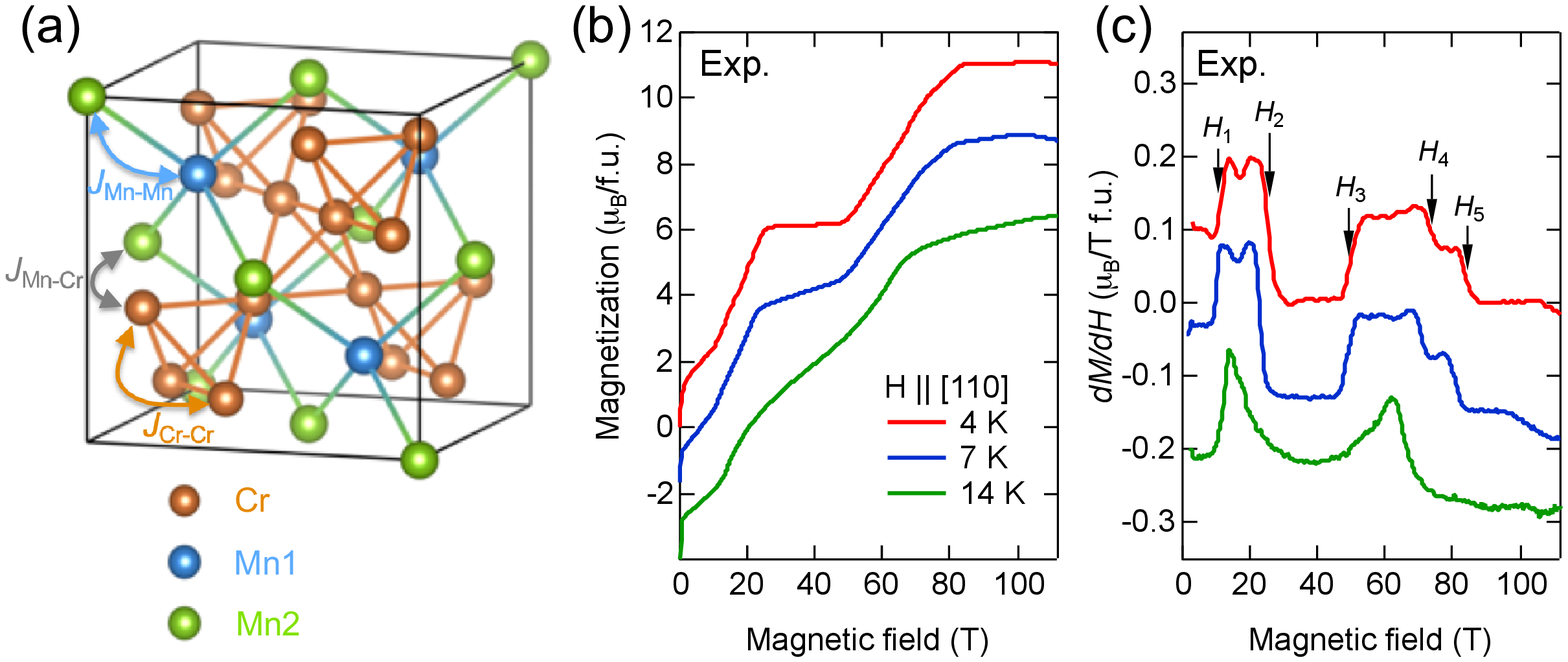}
----
\caption{(a) Crystal structure of MnCr$_2$S$_4$ and the three main magnetic interactions, \emph{J}$_\text{Mn-Mn}$, \emph{J}$_\text{Mn-Cr}$, and \emph{J}$_\text{Cr-Cr}$.
  (b) Magnetization and (c) its field derivative as a function of the magnetic field along the $[110]$ direction.
  The curves at 4~K are shown on the original scale, while the curves at 7 and 14 K are vertically shifted for clarity.
  In (c), the arrows indicate the phase transitions, $H_1$ -- $H_5$, at which the step-like anomalies corresponding to the sudden changes of the slope in the magnetization were observed.
  The magnetization saturates in the fully polarized state above 85~T.}

\end{figure*}

\section{Results}
\subsection{Magnetization measurements under ultrahigh magnetic fields}

We measured the magnetization $M$ of MnCr$_2$S$_4$ at 4, 7, and 14~K in pulsed fields up to 110~T shown in Fig.~1b.
In Fig.~1c, the field derivatives of the magnetization curves $dM/dH$ are presented. Focusing on the 4~K derivative in Fig.~1c, we easily identify five distinct steps labeled as $H_1$--$H_5$, which characterize field-induced transitions. 
The observation of an extremely robust magnetization plateau between 25 ($H_2$) and 50~T ($H_3$) is consistent with the previous study up to 60~T \cite{Tsu17}. 
From the magnetization plateau at 6~$\mu$$_\text{B}$/f.u. and the strong ferromagnetic interactions between the Cr spins \emph{J}$_\text{Cr-Cr}$, one can expect that the observed moment of 6~$\mu$$_\text{B}$/f.u. is attributed to the full ferromagnetic moment of the Cr spins and zero net moment of the antiparallel Mn spins.
The spin state below 10~T ($H_1$) corresponds to the well-known coplanar YK-type structure, where the two Mn-sublattice spins are canted and their net magnetization moment is antiparallel to the Cr spins.

To connect continuously from these canted Mn spins (in the YK phase) to the antiferromagnetically collinear Mn spins (in the plateau phase) without magnetization jumps, one can assume that the canted Mn spins rotate in the plane with changing the angle between two spins in the intermediate phase between $H_1$ and $H_2$.
As discussed in the following section, this asymmetric spin structure appearing in the intermediate phase is supported by classical Monte Carlo (MC) calculations (see Fig.~2d).

Above the plateau phase, we found two additional phase transitions at approximately 75 ($H_4$) and 85~T ($H_5$) in the ultrahigh-field experiments up to 110~T. 
Remarkably, the behavior observed below and above the plateau is symmetric: the plateau is sandwiched by two phases with steep slopes of the magnetization curve and then by two phases with more gentle slopes.  
The broadened $dM/dH$ data at 14 K in Fig.~1c also show a reasonably symmetry with respect to the magnetic field of 40~T. 
Based on this symmetry, one can assume that the two Mn spins rotate again through a spin-asymmetric phase between $H_3$ and $H_4$ and then a spin-canted phase appears between $H_4$ and $H_5$.
Finally, full polarization of 11~$\mu$$_\text{B}$/f.u. is found above $H_5$. Our MC calculations support this assumption as shown below.

\subsection{Spin-lattice coupling model}

\begin{figure*}[t]
----
\centering
\includegraphics[angle=0,width=0.8\textwidth]{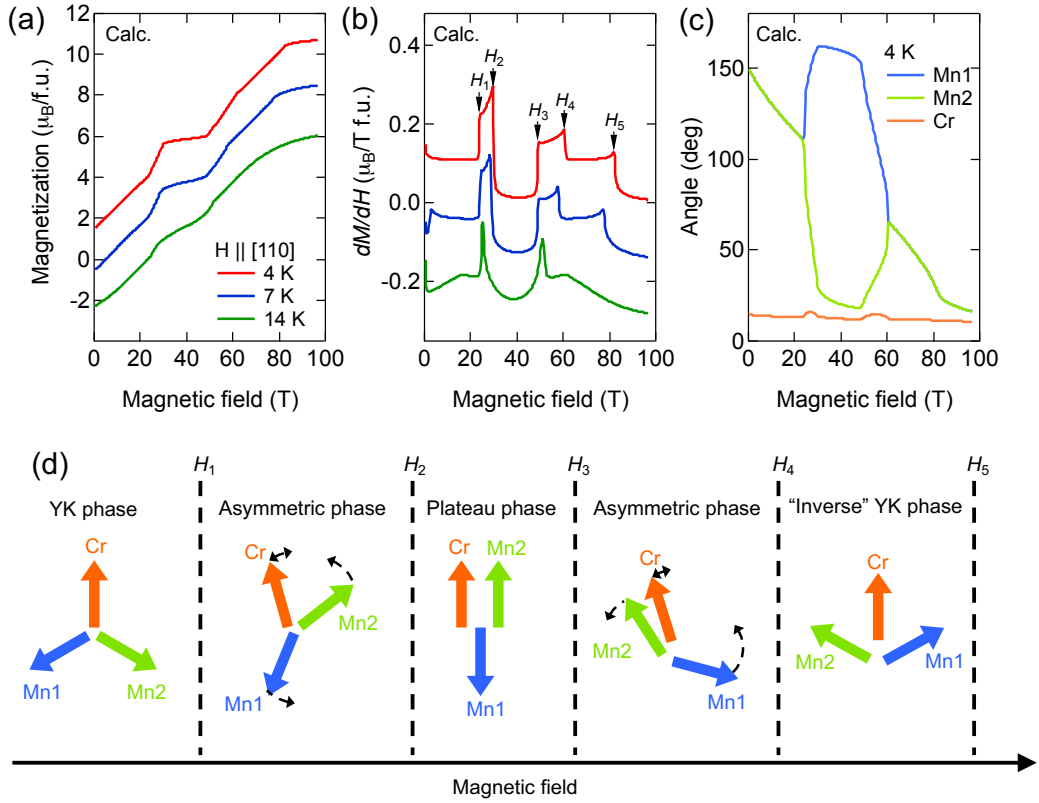}
----
 \caption{(a) Magnetization and (b) its field derivative as a function of the magnetic field applied along the $[110]$ direction calculated from classical Monte Carlo simulations. The curves at 4~K are shown on the original scale, while the curves at 7 and 14 K are vertically shifted for clarity. 
   The arrows ($H_1$ -- $H_5$) indicate the transitions similar to Fig.~1c. (c) Angles of the Cr, Mn1, and Mn2 spins with respect to the external-field direction.
   (d) Typical magnetic structures formed by Cr, Mn1, and Mn2 spins for each phase.}

\end{figure*}

Spin-lattice coupling is expected to play an essential role in magnetic properties of MnCr$_2$S$_4$. 
The previous ultrasound experiments showed strong anomalies in the velocity of the longitudinal sound waves at $H_1$, $H_2$, and $H_3$ \cite{Tsu17}. This observation indicates a significant spin-lattice coupling in the material.

Spin-lattice coupling has been discussed for antiferromagnetic chromium spinel oxides, CdCr$_2$O$_4$ \cite{Ued05, Miy13} and HgCr$_2$O$_4$ \cite{Ued06, Mat07}, where only Cr is the magnetic ion. Penc \textit{et al}. \cite{Pen04} theoretically predicted that a collinear spin configuration is favored and a robust magnetization plateau appears as a result of spin-lattice coupling, which introduces biquadratic spin interactions after tracing out the lattice degrees of freedom. In order to render the plateau robust for MnCr$_2$S$_4$, where both Mn and Cr ions are magnetic, it is reasonable to assume collinear Mn and Cr spins. We thus take into account a biquadratic term between Mn and Cr spins, $b_\text{Mn-Cr}\big({\bf S}_{\text{Mn}\textit{}}\cdot {\bf S}_{\text{Cr}\textit{}})^2$. 

Although the importance of the biquadratic terms in MnCr$_2$S$_4$ has been pointed out in the past~\cite{Nog79, Plu80}, this Mn-Cr biquadratic term has been neglected so far.

We propose the following minimal model:

\begin{eqnarray}
  \mathcal{H} = \mathcal{H}_\text{MM} + \mathcal{H}_\text{CC} + \mathcal{H}_\text{MC} + \mathcal{H}_\text{Z},
\end{eqnarray}
where
\begin{eqnarray}
  \mathcal{H}_\text{MM} &=& \sum_{\langle ij \rangle}  J_\text{Mn-Mn} \big({\bf S}_{\text{Mn}\textit{i}}\cdot {\bf S}_{\text{Mn}\textit{j}}) \nonumber , \\
  \mathcal{H}_\text{CC} &=& \sum_{\langle ij \rangle} J_\text{Cr-Cr} \big({\bf S}_{\text{Cr}\textit{i}}\cdot {\bf S}_{\text{Cr}\textit{j}}) \nonumber , \\
  \mathcal{H}_\text{MC} &=& \sum_{\langle ij \rangle} J_\text{Mn-Cr} (1-\alpha\rho_{ij}) \big({\bf S}_{\text{Mn}\textit{i}}\cdot {\bf S}_{\text{Cr}\textit{j}}) + \frac{K}{2}\rho^2_{ij} \nonumber ,\\
  \mathcal{H}_\text{Z} &=& - g\mu_\text{B}  {\bf H}\cdot(\sum_i {\bf S}_{\text{Mn}\textit{i}} + \sum_j {\bf S}_{\text{Cr}\textit{j}}). \nonumber
\end{eqnarray}
In this Hamiltonian, $\langle ij \rangle$ runs over the nearest neighbor within or between the Mn and Cr lattices, $\rho_{ij}$ describes the lattice displacement between neighboring Mn and Cr ions, $K$ and $\alpha$ are the spring constant and the spin-lattice coupling, $g \approx 2$ and $\mu_\text{B}$ are the g-factor and the Bohr magneton, respectively.

We expect the interaction between Mn and Cr spins to be more sensitive to the positions of atoms than that between Mn spins. It is because the superexchange path of a Mn-Cr bond is involved with a single sulfur atom, while the path of a Mn-Mn bond is involved with multiple sulfur atoms. The angle formed by Mn-S-Cr atoms, which is naturally parameterized by the bond phonon, should affect the interaction between Mn and Cr spins significantly. It is, therefore, reasonable to include the spin-lattice coupling only in the Mn-Cr spin interaction as a minimal model.

After tracing out $\rho_{ij}$, we can exactly rewrite the interactions between the Mn and Cr lattices as
\begin{eqnarray}
  \mathcal{H}_\text{MC} =  \sum_{\langle ij \rangle} J_\text{Mn-Cr} \big({\bf S}_{\text{Mn}\textit{i}}\cdot {\bf S}_{\text{Cr}\textit{j}}) - b_\text{Mn-Cr}\big({\bf S}_{\text{Mn}\textit{i}}\cdot {\bf S}_{\text{Cr}\textit{j}})^2 \nonumber ,
\end{eqnarray}
where the biquadratic term is introduced with the coefficient $b_\text{Mn-Cr}=J^2_\text{Mn-Cr}\alpha^2$/$2K$, which we use as a parameter for the spin-lattice coupling below. Despite the integrability of lattice displacements, we intentionally retain them in the model~(1) to compare theoretical results to experimental data regarding not only \textit{spin} (magnetization) but also \textit{lattice} (magnetostriction and sound velocity) degrees of freedom.

Note that the magneto-crystalline anisotropy of MnCr$_2$S$_4$ is weak \cite{Tsu02}, because Cr ions have a half-filled $t_{2g}$ shell  (3$d^3$) and Mn ions have a half-filled $t_{2g}$ and $e_g$ shell (3$d^5$) with zero orbital momentum.

\subsection{Classical Monte Carlo calculations: Magnetization and spin angles}

\begin{figure*}[t]
\centering
\includegraphics[width=0.8\textwidth]{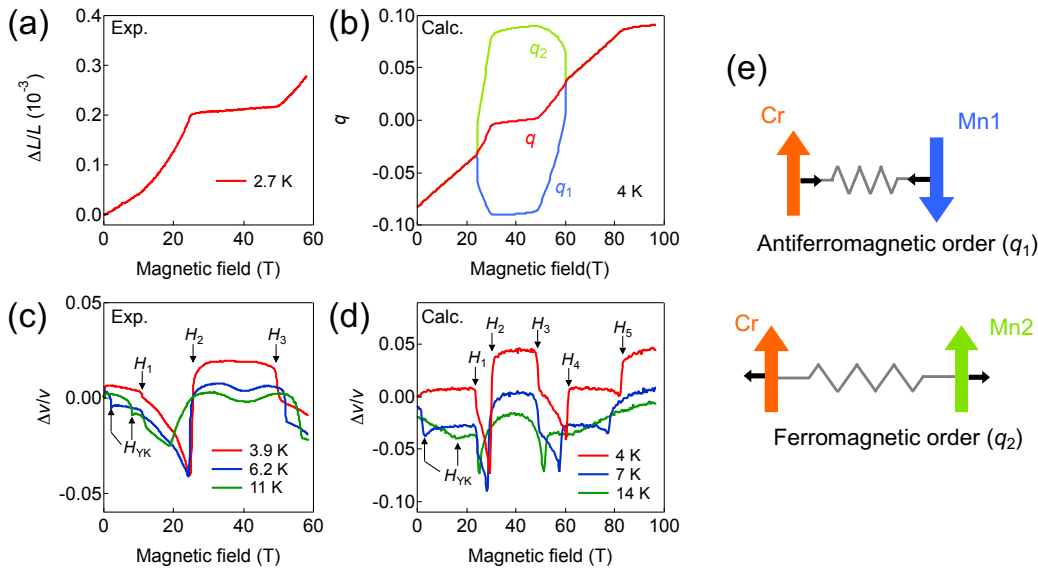}
 \caption{
Magnetostriction and sound velocity obtained by experiment and theory.
(a) Experimentally obtained magnetostriction along $\bf{H}$ $\parallel [110]$.
(b) Exchange-modification parameters $q_1$ and $q_2$, which are proportional to the Cr-Mn1 and Cr-Mn2 bond lengths, respectively. The magnetostriction measurement $\Delta L/L$ should be proportional to $q=(q_1+q_2)/2$. In (b), $q_1$ and $q_2$ are identical below 20~T and above 60~T.
(c, d) Relative changes of the sound velocity obtained in experiment and theory.
The longitudinal acoustic mode $(c_{11}+2c_{12}+4c_{44})/3$ propagating along  $\bf{H}$ $\parallel [111]$ was measured.
(e) Schematic illustration of antiferromagnetically ordered spins with $q_1$ and ferromagnetically ordered spins with $q_2$ in the plateau phase. 
}
\end{figure*}

We performed classical Monte Carlo simulations for the model~(1) and calculated the magnetization, $M$, and the field derivatives of the magnetization, $dM/dH$, at 4, 7, and 14~K as shown in Figs.~2a and 2b. Indeed, our simple model already accounts for the experimental data (Figs.~1b and 1c). The five distinct steps, $H_1$--$H_5$, are all reproduced at 4~K. The two intermediate phases from $H_1$ to $H_2$ and from $H_3$ to $H_4$, which are adjacent to the magnetization plateau, have the steep slopes of the magnetization curve in a similar way as the experimental data.

We optimized the parameters in the model, \emph{J}$_\text{Mn-Mn}$=~3.4 K, \emph{J}$_\text{Mn-Cr}$=~3.1 K, \emph{J}$_\text{Cr-Cr}$ =~-9.1 K, and \emph{b}$_\text{Mn-Cr}$ = 0.04 K, by fitting the calculated magnetization to the experimental data. Some parameters were simultaneously optimized to reproduce the two transitions at $T_\text{C}\approx$ 65~K and at $T_\text{YK}\approx$ 5 K observed in previous experiments without magnetic field~\cite{Lot56, Tsu03} (see Appendix A). Our estimates differ only by a few percent from the previous estimates~\cite{Tsu17}, $J_\text{Mn-Mn}=3.1$~K and $J_\text{Mn-Cr}=3.24$~K ~\cite{J_value}. In our spin-lattice model, the effective exchange coupling between Mn and Cr spins becomes $ J_\text{Mn-Cr} (1-\alpha\rho_{ij}) \approx J_\text{Mn-Mn}$ around zero magnetic field, which is consistent with the YK model~\cite{Lot68}. The internal magnetic field that is created by the Cr spins and acting on the Mn spins is estimated to  be about 42~T using \emph{J}$_\text{Mn-Cr}$=  3.1 K. This is consistent with the observation of the symmetric behavior with respect to the field of 40~T.

From the quantitative point of view, the two intermediate phases obtained theoretically are narrower than those observed experimentally. We incorporated additional biquadratic terms, $b_\text{Mn-Mn}\big({\bf S}_{\text{Mn}}\cdot {\bf S}_{\text{Mn}})^2$, but they do not improve the results (not shown here). A more complex model is required for a quantitative comparison to the experimental data.

To investigate the magnetic structures of the nontrivial phases induced by magnetic field, we calculated the angles of the Cr and Mn spins from the external-field axis at 4~K, shown in Fig.~2c.
Using this data, we sketched typical magnetic structures for each phase in Fig.~2d.  The spin-lattice coupling stabilizes not only the magnetization plateau with a collinear magnetic structure but also the two intermediate phases, where a triangular-like structure formed by one Cr and two Mn spins rotates continuously with magnetic fields variation. The intermediate phases give rise to unconventional asymmetric magnetic structures.

\subsection{Magnetostriction and sound velocity}

MC calculations for the model~(1) account for the magnetostriction and the sound velocity as well.
The experimentally obtained magnetostriction $\Delta L/L$ is shown in Fig.~3a.
Figure~3b shows the exchange-modification parameters $q_1=\alpha \rho_\mathrm{Mn1-Cr}$, $q_2=\alpha \rho_\mathrm{Mn2-Cr}$, and $q=(q_1+q_2)/2$ obtained by the MC calculations.

The averaged displacement $q$ is proportional to the change of the sample length, which is measured in the magnetostriction experiment.

The relative change of the sound velocity $\Delta v/v$ obtained experimentally and theoretically are shown in Figs. 3c and 3d, respectively.
In the MC calculations, we deduced the change of the sound velocity from the effective spring constant $K_\text{eff}$, which differs from the original constant $K$ owing to the spin-lattice coupling.
We calculated the compressibility $\kappa$ in the MC simulations: $\kappa \equiv \beta \text{Var}[q] \sim \beta \int dq q^2 e^{-\frac{\beta K_\text{eff}}{2}q^2} / \int dq e^{-\frac{\beta K_\text{eff}}{2}q^2} = 1/K_\text{eff}$, where $\beta$ is the inverse temperature.
Although acoustic phonon modes are not included in our model~(1), the effective spring constant should be common to various phonon modes. The sound velocity measured in experiments should be related to the effective spring constant: $v \propto \sqrt{K_\text{eff}}$. Therefore, we can estimate the velocity using the relation $v \propto 1/\sqrt{\kappa}$.

Our theoretical model accounts for all the characteristic features related to the phase transitions ($H_1$--$H_3$) observed in the experiments up to 60 T.
Moreover, the low-field anomalies labelled as $H_\text{YK}$ (2 T at 6.2 K and 9 T at 11 K in Fig.~3c), which corresponds to a phase boundary between the YK phase and the ferrimagnetic phase, where only the Cr spins order~\cite{Tsu03}, are also reproduced in the MC simulations as shown in Figs.~3c and 3d. The agreement between experiment and theory clearly shows the validity of our model and the relevance of the magnetoelastic coupling between the Mn and Cr ions.

The theory reproduces the magnetostriction data very well and suggests that the spin-lattice coupling produces strong ($q_1<0$) and weak ($q_2>0$) bonds between the Mn and Cr spins (see sketch in Fig.~3e). In the plateau phase, the Cr and Mn spins are antiferromagnetically (ferromagnetically) ordered on the strong (weak) bonds.
In other words, this $Z_2$ symmetry of the Mn-Cr bond is spontaneously broken. Thus, a domain-wall excitation related to the $Z_2$ symmetry is expected to appear at some finite temperatures due to the entropic effect.
The broad minimum observed around 40~T in the ultrasound data at higher temperatures (6.2 K and 11 K), which is absent in the theoretical results, might be attributed to dynamics of the domain wall.

\subsection{Phase diagram}

Combining the present ultrahigh magnetic-field experiments with previous data in ref. \cite{Tsu17}, we obtain the complete phase diagram of MnCr$_2$S$_4$ in Fig.~4. The magnetic structures for all phases, identified by our MC calculations, are schematically shown in the figure. 
To identify the higher temperature phase above $T_\text{YK}$ at zero magnetic field, we calculated the spin correlations between the Mn1 and Mn2 spins using the model~(1) with the same parameters (see Appendix B). 
While the correlation between the Mn spin components parallel to the Cr spins develops below $T_\text{C}$, the perpendicular components are almost independent down to $T_\text{YK}$. This indicates that the transverse components of the Mn spins remain paramagnetic in the intermediate temperature phase between $T_\text{C}$ and $T_\text{YK}$, and the antiferromagnetic order of the transverse components eventually takes place at $T_\text{YK}$ as shown in Fig.~4.

With increasing temperatures, both spin-asymmetric phases diminish on the expense of the expanding spin-collinear phase. 
This finding indicates that thermal fluctuations favor the collinear magnetic structure rather than noncollinear ones \cite{Kaw85, Zhi02}. 
In the phase diagram, the higher-field phases mirror the lower-field phases; the phase diagram is symmetric with respect to approximately 40~T, where the external fields cancel out the internal exchange fields acting on the Mn spins.

\begin{figure}[tbh]
\centering
\includegraphics[width=0.48\textwidth]{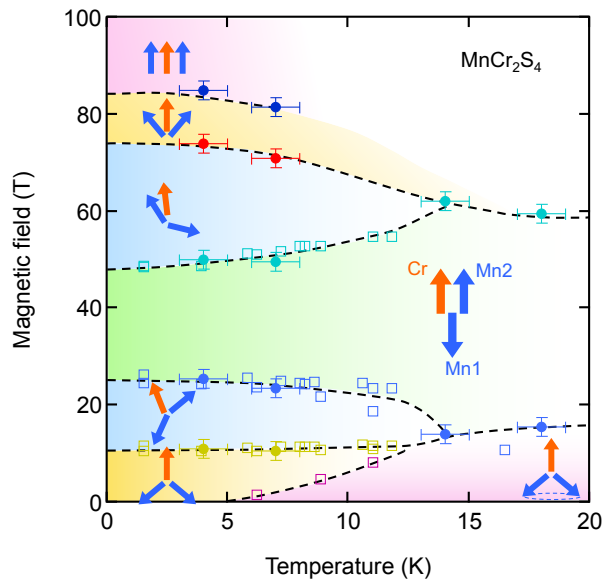}
 \caption{
Magnetic-field-temperature phase diagram of MnCr$_2$S$_4$. The phase boundaries were obtained from the magnetization, ultrasound propagation and magnetostriction experiments in pulsed fields. The closed circles with error bars of $\pm2$~T and $\pm1$~K were obtained in the present ultrahigh-field experiments. The open squares were taken from
ref. \cite{Tsu17}. Typical magnetic structures revealed by the MC calculations are illustrated in each phase. The Cr spins (orange arrows) are almost perfectly aligned with the external field. The Mn spins (two blue arrows) show a complex, strongly field-dependent order.  
}
\end{figure}

\section{Discussions}
Thanks to our measurements in ultrahigh magnetic fields, we unraveled the \textit{complete} $H$--$T$ phase diagram of the YK ferrimagnetic spinel compound. The experimental results are accounted for by a quite simple model incorporating spin-lattice coupling between the Mn and Cr ions.
Our findings indicate that the spin-lattice coupling is crucial for establishing the extremely robust magnetization plateau.
The effect of spin-lattice couplings has been extensively investigated both experimentally and theoretically in antiferromagnetic Cr spinel oxides, ZnCr$_2$O$_4$ \cite{Miy11}, CdCr$_2$O$_4$ \cite{Ued05}, and HgCr$_2$O$_4$ \cite{Mat07}, where the pyrochlore lattice reveals strong geometrical frustration.
Here, with only weak frustration caused by the competition between \emph{J}$_\text{Mn-Mn}$ and \emph{J}$_\text{Mn-Cr}$ and  with two different magnetic ions, Mn and Cr, still a robust magnetization plateau was observed.
Our spin-lattice model appears to be universally applicable to a broad range of \textit{ferrimagnetic} and \textit{antiferromagnetic} spinel compounds.
In fact, strong spin-lattice couplings have been widely expected to exist in spinel compounds, such as in CoCr$_2$O$_4$ \cite{Tsu13} and ZnCr$_2$S$_4$ \cite{Fel12}, where unconventional magneto-structural phase transitions have been observed. They might be understood incorporating spin-lattice couplings to a simple spin model in a similar way as done in the present study.

Another important finding is that the unconventional triangular-like magnetic structures in the two intermediate phases are formed by one Cr and two Mn spins rotating continuously as magnetic fields increase. 
This identification finally gives a clear solution to the long-standing problem of the high-field magnetic structures of MnCr$_2$S$_4$ \cite{Nog79, Plu80, Tsu17, Den70}. 
The origin of the intermediate phases is the competition between the YK-type interactions, which favors a \textit{noncollinear} triangular structure, and the spin-lattice coupling, which favors a \textit{collinear} structure.
Here, we gain the straightforward insight that incorporation of spin-lattice couplings in ferrimagnetic spinel compounds leads to a variety of unconventional magnetic phases, such as fractional magnetization plateaus and spin-driven multiferroic phases.

Lastly, we discuss possible improvements in our spin-lattice model.
In the ultrasound experimental data (Figs.~3c and d), anomalous features were observed around the middle of the magnetization plateau at higher temperatures, which is absent in our theoretical consideration. This anomaly is even more pronounced in the ultrasound attenuation as shown in previous experiments \cite{Tsu17}. 
It was speculated that this anomalous behavior might be attributed to a frustration in the Mn diamond lattice induced by next-nearest-neighbour interactions, as recently discussed for another spinel compound, MnSc$_2$S$_4$ \cite{Ber07, Gao17}.
The inclusion of farther spin interactions to our minimal model~(1) might be appropriate.

Another possibility to improve the agreement may lie in more complex interactions of the lattice degrees of freedom. In our minimal model, lattice displacements are independent of each other, forming a flat phonon band. A phonon band structure formed by connected lattice degrees of freedom could more accurately describe relevant lattice displacements and reproduce the anomaly observed in experiment.

Although it has been believed for a long time that a majority of magnetic properties of ferrimagnetic spinels can be understood within the YK and LKDM models, our studies have revealed that additional perturbations, such as the spin-lattice coupling, are crucial to explain their properties under magnetic fields. Incorporating spin-lattice coupling, the YK and LKDM models reopen a new platform to study unconventional magnetic phases, such as fractional magnetization plateaus and spin-driven multiferroic phases.

\section{Acknowledgements}

This research has been supported by the DFG via TRR 80 (Augsburg - Munich), by SFB 1143 (Dresden), through the W\"urzburg-Dresden Cluster of Excellence on Complexity and Topology in Quantum Matter - \textit{ct.qmat} (EXC 2147, project-id 39085490), and by the BMBF via DAAD (project-id 57457940). We acknowledge support by the project 16.80012.02.03F (ASM) and by HLD at HZDR and LNCMI at CNRS, both members of the European Magnetic Field Laboratory (EMFL).

\section{Appendix A: Magnetic susceptibility and sound velocity}

We calculated the inverse magnetic susceptibility $\chi^{-1}$ and the relative change of the sound velocity $\Delta v/v$ at temperatures ranging from 2.5 to 400 K using the classical Monte Carlo simulation for the model (1) in the main text. The parameters were set to the same values as those in the main text: \emph{J}$_\text{Mn-Mn}$ = 3.4~K, \emph{J}$_\text{Mn-Cr}$ = 3.1~K, \emph{J}$_\text{Cr-Cr}$ = -9.1~K, and \emph{b}$_\text{Mn-Cr}$ = \emph{J}$^2_\text{Mn-Cr}\alpha^2/2K$ = 0.04~K. The quantum-classical crossover of the spin system needs to be correctly considered~\cite{Hub08}: We treat the spins as vectors and set the spin (vector) lengths to $|{\bf S}|_\text{cl}=S$ below 15~K and $|{\bf S}|_\text{cl}=\sqrt{S(S+1)}$ above 20~K.
The classical Monte Carlo simulations qualitatively reproduce experimental data observed by Tsurkan \textit{et al}. in ref.~\cite{Tsu17}.

\begin{figure}[t]
----
\centering
\includegraphics[angle=0,width=0.48\textwidth]{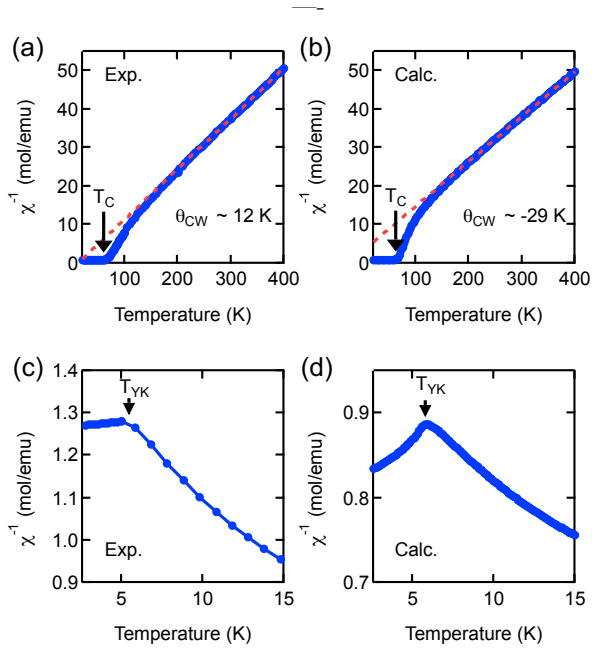}
----
 \caption{Inverse magnetic susceptibility $\chi^{-1}$ of MnCr$_2$S$_4$ obtained in the experiments~\cite{Tsu17} and the theory. An external magnetic field of 1 T is applied along the $[111]$ direction. (a) and (b) show the data at high temperatures of 20--400 K. (c) and (d) show the data at low temperatures of 2.5--15 K.}

\end{figure}

\begin{figure}[t]
----
\centering
\includegraphics[angle=0,width=0.48\textwidth]{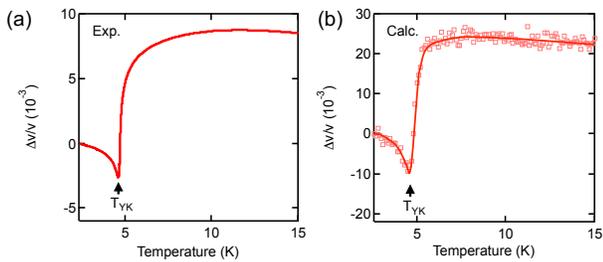}
----
 \caption{Relative changes of elastic constant $\Delta c/c$ of MnCr$_2$S$_4$ obtained in the experiments~(a) ~\cite{Tsu17} and the theory~(b).}

\end{figure}

\section{Appendix B: Spin correlation}
To investigate a magnetic phase between T$_\text{YK}$ and T$_\text{C}$, we calculated spin correlations of the Mn1 and Mn2 spins, S$_\text{Mn1}^{\parallel}$S$_\text{Mn2}^{\parallel}$ (parallel component along the external field) and S$_\text{Mn1}^{\perp}$S$_\text{Mn2}^{\perp}$ (perpendicular component against the external field).

\begin{figure}[h]
----
\centering
\includegraphics[angle=0,width=0.48\textwidth]{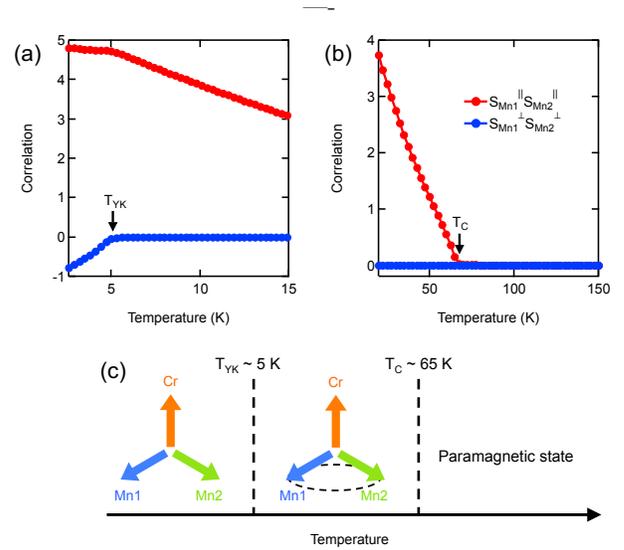}
----
 \caption{Spin correlations of the Mn1 and Mn2 spins at low temperatures~(a) and high temperatures~(b). (c) shows schematic magnetic structures of the Yafet-Kittel triangular-structure phase (T $<$ T$_\text{YK}$) and the ferrimagnetic phase with disordered transverse components of Mn spins (T$_\text{YK}$ $<$ T $<$ T$_\text{C}$).}

\end{figure}

\end{document}